# Intelligent Management and Efficient Operation of Big Data

**José Moura[1,3], Fernando Batista[1,2], Elsa Cardoso[1,2], Luís Nunes[1,3]**
[1] *ISCTE-IUL, Instituto Universitário de Lisboa, Portugal*
[2] *INESC-ID, Lisboa, Portugal*
[3] *IT, Instituto de Telecomunicações, Lisboa, Portugal*
{jose.moura, fernando.batista, elsa.cardoso, luis.nunes}@iscte.pt

**ABSTRACT**

*This chapter details how Big Data can be used and implemented in networking and computing infrastructures. Specifically, it addresses three main aspects: the timely extraction of relevant knowledge from heterogeneous, and very often unstructured large data sources; the enhancement on the performance of processing and networking (cloud) infrastructures that are the most important foundational pillars of Big Data applications or services; and novel ways to efficiently manage network infrastructures with high-level composed policies for supporting the transmission of large amounts of data with distinct requisites (video vs. non-video). A case study involving an intelligent management solution to route data traffic with diverse requirements in a wide area Internet Exchange Point is presented, discussed in the context of Big Data, and evaluated.*

## 1. INTRODUCTION

Big Data is a relatively new concept. When someone is asked to define it, the tale of the blind man and the elephant immediately comes to mind. As in the tale, each person that talks about Big Data seems to have his/her own view, according to the person's background or the intended use of the data (Ward & Barker, 2013; McAfee & Brynjolfsson, 2012; Cox & Ellsworth, 1997; Diebold, 2012; Press, 2013). Big Data is closely related to the area of analytics (Davenport et al., 2010), as it also seeks to gather intelligence from data generating value to the business or the organization. However, a Big Data application differs in terms of the volume (referring to large data volumes), velocity (i.e., multi-structured data types) and variety (related to the change rate and time-sensitive usage to maximize the business value) of the data involved. These aspects are usually known as the 3V's. These large and diverse data streams require "ever-increasing processing speeds, yet must be stored economically and fed back into business-process life cycles in a timely manner," (Michael & Miller, 2013, pp. 22). Big Data applications offer new opportunities of information processing for enhanced insight and decision-making in different disciplines such as business, finance, healthcare, transportation, research, and politics.

The successful deployment of a Big Data infrastructure requires the extraction of relevant knowledge from original heterogeneous (Parise et al, 2012), highly complex (Nature, 2008) and massive amount of data. To this end, several tools from different areas can be applied: Business Intelligence (BI) and Online Analytical Processing (OLAP), Cluster Analysis, Crowdsourcing, Network Analysis, Text Mining, and Natural Language Processing (NLP). As an example, massive amounts of textual information are constantly being produced and can be accessed from online sources, including social networks, blogs, and numerous websites. Such unstructured texts represent potentially valuable knowledge for companies, organizations, and governments. The process of extracting useful information from such unstructured texts, known as Text Mining, is now becoming a relevant research area. It draws from different fields of computer science, such as Web Mining, Information Retrieval (IR), NLP, Machine Learning (ML), and Data Mining. Today's text mining research and technology enables high-performance analytics from web's textual data, allowing to: cluster documents and web pages according to their content, find associations among entities (people, places and/or organizations), and reasoning about important data trends.

The data sets in Big Data are becoming increasingly complex (Nature, 2008). For example, the biology field is urging for robust data computing (The Apache Software Foundation, 2014a) and distributed storage solutions (The Apache Software Foundation, 2014c); machine learning algorithms for data mining tasks (Hall et al., 2009); online community collaborations need wiki-style information cooperative tools (Waldrop, 2008); sophisticated visualization techniques of intracellular signaling pathways require tools like GenMAPP (Waldrop, 2008); and innovative ways to control the Big Data infrastructure such as software-design networking (SDN). To conclude, Lawrence Hunter, a biological researcher, wrote: "Getting the most from the data requires interpreting them in light of all the relevant prior knowledge," (Marx, 2013). Clearly, satisfying this requisite also demands for new scalable Big Data solutions. In this way, the Big Data is a very challenging and exciting research area to be further explored and investigated.

An important aspect to guarantee the success of Big Data solutions is to manage with more intelligence the supporting computing/networking infrastructure. Currently, both data and collaborative applications are increasingly being moved towards Data Centers aggregated inside the cloud. Consequently, to obtain a good performance in Big Data applications it is mandatory to achieve a proper performance in Data Centers. To achieve this, some management enhancements are needed to operate more intelligently the available resources, such as: virtual machines (Dai et al., 2013), memory (Zhou & Li, 2013), CPU scheduling (Bae et al., 2012), cache (Koller et al., 2011), I/O (Ram et al., 2013), and network (Marx, 2013; Lange et al., 2011; Saleem, Hassan, & Asirvadam, 2011).

This chapter details how Big Data can be deployed and used, focusing on the following important aspects: i) timely extraction of relevant knowledge from heterogeneous, and often unstructured data sources; i) enhancement of the performance of processing and networking (cloud) infrastructures for Big Data, using more intelligent management solutions/algorithms; and iii) SDN as an intelligent and very interesting solution to conveniently manage the network infrastructures to transmit Big Data with diverse functional requisites. At the end of this chapter, a SDN application is discussed in a wide area network (WAN) environment, more specifically at an Internet exchange point (IXP). This case study shows that using high-level and intuitive management policies is possible to easily route data traffic with diverse requirements among network infrastructures owned by several entities, which is a very realistic scenario to deploy Big Data services through diverse public cloud providers, supporting the following three Big Data aspects: dynamic, diversity and distributed.

## 2. FROM BIG DATA TO USABLE KNOWLEDGE

The heuristics "There is no Data Like More Data" is famous in NLP and was made back in 1985, even before the web (Jelinek, 2005). A few years later, the web and the increasingly amounts of available content supported this heuristic even further. Suddenly, all the investment went to data and statistical methods research, supported by the widely spread idea that data would be the only important thing. This feeling was reflected precisely on the famous quote "Every time I fire a linguist, the performance of the speech recognizer goes up" attributed to Frederick Jelinek and also on the title "Every time I fire a linguist, my performance goes up" used by Hirschberg (1998) for one of her talks. In more recent years, the most revelant challenges in terms of data annotation have been on: how to produce annotated data with a minimal supervision; on the use of co-training, a semi-supervised learning approach based on two different aspects of the data; and on the use of bootstrapping, which consists of calculating the initial parameters from related information sources.

The challenge of big data is not, as in previous decades, to gather it, but to find a way to make use of it. In the last decades hundreds of machine learning algorithms were developed, but testing them was often difficult. Data was scarce and the algorithms that promised good scaling qualities always had a difficult time proving it. Today several companies give away huge data sets to run tests (Netflix, Amazon, Kaggle) and others (Twitter) make public data-streams available. Many efforts have been made during this last decade in finding the best methodology to tackle the increasingly amounts of available data. As a result, the incredible advances in ML methods and related technologies make it possible to continue strengthening the focus on the data. The main issue is what to do with all that data? From a Machine Learning point of view, more data usually means more reliable models and results, but also more processing time. Some ML methods may become unusable due to the amount of data, others may require large numbers of parameters, but some may even become simpler, due to the clear definition of the desired patterns and the outliers. In terms of tools, many are now surfacing, but also all major ML and Data Mining tools are upgrading to deal with the new problems. In the next sections we will review some of the most important Big Data problems (e.g. scale, speed, and data types), the ML methods used and the new role of data scientist.

## Big challenges

Big challenges differ from others mainly in scale, speed and types of data used. Nevertheless, the underlying learning mechanisms are still similar, mainly because researchers in the ML field have been dealing with large amounts of data, continuous data series, unstructured data, and focusing on scalability problems, for quite some time. Speech processing for example has always been a Big Data problem, although some years ago the problem started when trying to gather data and store it and label it. Now, most of these difficulties are solved for the most common languages with the large, publicly available, corpora and large data-storage devices. Following, we discuss how to apply ML to Big Data analysis.

On one hand, from a ML point of view, there are three main types of learning systems: Supervised, Unsupervised and Reinforced. The first type requires examples of the correct behavior or classification, the second requires only examples and (often) a distance function between examples, and the third a function to evaluate the system state. It is also possible to join search algorithms to these three broad categories.

On the other hand, the main types of analysis problems addressed by Big Data tools are the following ones:

- Clustering
- Classification
- Recommending systems
- Frequent item set mining

We now briefly describe each type of Big Data analysis problem, using the more suitable set of ML solutions, and giving a corresponding illustrative application example. Clustering is a type of unsupervised learning where the goal is to divide the data into associated groups (clusters). A typical example is learning social circles, and patterns of how information is spread.

Classification is a supervised learning problem, which consists on learning a model of a system that performs a given classification based on correct (labeled) classification examples. An example is credit card fraud analysis. It is possible to train a model with information of good and bad transactions and require that the system, given an unseen situation, provide the correct classification (with a given error margin).

Recommending systems are labeled as collaborative filtering and can also be seen as a form of supervised learning, although unsupervised learning methods are often used to compose a solution for problems of this type. The problem consists on figuring what is the best product to recommend to a user, given its own history and the histories of all other users. Amazon has been using recommending systems since the 90's to suggest products to returning costumers (Linden, Smith, & York, 2003).

Frequent item-set mining is also a form of Association Rule Learning and Unsupervised Learning. The typical problem is market basket analysis that tries to detect shopping patterns by looking for sets of items that repeat frequently.

Often, the problem is not only that the sets themselves are huge is that data is an endless continuous stream (Twitter, Facebook, Netflix), or comes in long multivariate sequences (data produced by a Kinect device during a game). Processing these sequences / streams in time for using the information is also a major issue.

To deal with this there are several strategies, depending on the problem and on the learning algorithm used:

- Don't use all the data at once: the idea of having a training-set and a test-set running based on your PC may need to be revised and your applications may need to pull data chunks from a data-set server and throw them away when used. Works best with stochastic or one-time algorithms that require less (or no) pattern replay.
- Use incremental algorithms that continue to learn as data comes in but do not require all previous examples to be stored.
- Pre-process to simplify data and use "the interesting bits". Several algorithms are known to discover highly correlated data and related attributes; the use of these can reduce the amount of information that actually needs to be processed. Don't underestimate human ability to find patterns, especially if equipped with a good visualization tool.
- Distribute the data and processing: The use of distributed file systems seems to be one where tools are making the most difference. Currently, tools that let you distribute your data in the cloud are the basis of the main Big Data approaches.

From the clutter of "Big Data tools" that are appearing by the dozens, both from upstarts as well as from established companies one name pops up, Hadoop Distributed File System (HDFS). Hadoop is a distributed computing platform based on its own file system HDFS. This seems to be the standard that most tools aim at supporting. The Apache Mahout Project (The Apache Software Foundation, 2014b) is a leading effort for open-source, scalable machine learning algorithms. Although still in its early years it already gathered some attention and a growing community. There are a multitude of tools, either adding scalable components, or being built from scratch for this new paradigm. However, contrary to most other revolutions in the area, tools do not seem to be the only thing changing. To cope with these "big data" problems organizations are now forced to value the knowledge that can be gained by good practitioners. Professionals with these skills are becoming highly valued and a new discipline is emerging, that was labeled Data Science.

## Data science: a new way of looking at information extraction from data

Organizations now begin to perceive the added value of having a team dedicated to data analysis supporting the decision making process. Analysts can be defined "as workers who use statistics, rigorous quantitative or qualitative analysis, and information modeling techniques to shape and make business decisions," (Davenport, Harris, & Morison, 2010). An effective analysis should impact the decision-making process. To do so, an analysis must be driven by a business need and be relevant to that business, i.e., providing insight to solving a relevant problem of the organization.

Currently, more than data analysts, organizations are seeking for data scientists to process, interpret and, mainly, find new ways of using the available data, regardless if it is internal or external, structured or un-structured, big or small. This emergent role can be characterized by a solid foundation in math, statistics, probability, and computer science, combined with strong social skills. Apart from the ability to write code and to communicate to all stakeholders (e.g., using data storytelling techniques), data scientists need to have an intense curiosity, leading him/her to dive into the root of a problem and derive a set of hypothesis that can be tested (Davenport & Patil, 2012). Given the nature of this skill set, the role of data scientist is still scarce in the marketplace.

Data Science is then an emerging discipline that encompasses data analysis, data mining, statistics, visualization, machine learning, and sometimes even business and marketing-related subjects.

## Text mining

Text Mining is the non-trivial process of extracting relevant knowledge from large collections of unstructured text documents. In this information age we are living, a relevant part of the information is available as unstructured texts, very difficult to access and to process, but still a potential source of useful knowledge.

Text mining is a recent interdisciplinary field of computer science that combines techniques from Natural Language Processing, Computational Linguistics, Data mining, Web mining, Information Retrieval and Statistics, and Machine Learning. Text Mining is about finding patterns on large unstructured text databases, contrarily to Data Mining that deals normally with large structured databases. Text Mining is also much different than traditional web search, where the user is typically looking for something that is already known and has been produced by someone else. While Text Mining focuses on extracting new information from text, traditional web search relies on filtering the material that is not currently relevant to the user's needs. Text Mining is also different from traditional Information Extraction, a major component of Text Mining that can be used to extract other high level information.

Today's text mining research and technology enables high-performance analytics from web's textual data, allowing to: cluster documents and web pages according to their content, find associations among entities (people, places and/or organizations), and reasoning about important data trends. General examples of existing Text Mining applications include: 1) classification of news stories, web pages, etc., according to their content; 2) email and news filtering; 3) organizing repositories of document-related meta-information for search and retrieval (search engines); 4) clustering documents or web pages; 5) gain insights about trends, relations between people, places and/or organizations. Text Mining applications are particularly relevant for different business aspects, including decision support in Customer Relationship Management, Marketing, and Industry. Nowadays companies are relying on text mining to discover and analyze customer's typical complaints, trends of satisfied and unsatisfied customers, and distinct groups of potential buyers, groups of competitors.

Despite the undeniable advances in the scope of Text Mining, a wide range of challenges still pose difficulties for Text Mining applications. One of the challenges relates to the fact that the number of possible words and phrase types in language cause a very high number of possible dimensions, apart from being sparse. Another important issue concerns the documents being sometimes structurally different and not statistically independent. Aspects related with the language complexity, subtle ways of establishing relations between concepts in a text, word ambiguity, and context sensitivity also pose considerable challenges. Finally, noisy data, such as spelling mistakes, and specific phenomena that is often found in microblogs, such as Twitter also pose considerable challenges to Text Mining applications.

Text Classification is a particularly important type of Text Mining tasks that consist of assigning a predefined set of classes (subject categories, topics, genres, sentiment, etc.) to a document. Examples of classification tasks include: Spam detection, Language Identification, Age/gender identification, Sentiment Analysis, Topic detection, and other tasks that are illustrated in Figure 1. Each one of the referred tasks makes use of NLP, statistics, and machine learning techniques to extract potentially relevant knowledge from the content.

Sentiment Analysis is a relevant and well-known task that consists of extracting sentiments and emotions expressed in texts. Being the first step towards the online reputation analysis, it is now gaining particular relevance because of the rise of social media, such as blogs and social networks. The increasing amount of user-generated contents in the form of reviews, recommendations, ratings and any other form of opinion constitute huge volumes of opinionated texts all over the web that are precious sources of information, especially for decision support. Sentiment Analysis can be used to know what people think about a product, a company, an event, or a political candidate. Agencies can make use of it to check how the public sentiment is, predict election outcomes, and even to predict market trends (Bollen, Mao, & Zeng, 2011).

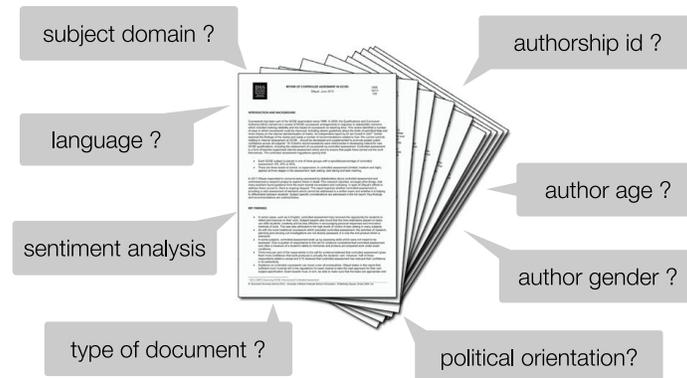

**Figure 1 -** Text classification tasks.

## Open issues

The challenges in this area are now beginning to be tapped. Most tools to analyze data are already available. Strategies to disperse data and processing in the cloud, while maintaining data privacy are also available and more are being developed.

The Data Science teams are starting to be considered a necessary asset for some companies, mainly for those whose business relies mainly on data. Applications seem to be everywhere, some are problems that have been around for a while and joined the Big Data stream; others appeared to be unsolvable until recently and are now being tackled for the first time; and others still have jumped to a new scale.

The incredible advances in Machine Learning methods, computational frameworks, and related technologies make it now possible to tackle immense quantities of data. However, learning with such data remains still a serious challenge. Most of the existing machine learning techniques works well because the process relies just on optimizing weights, based of human-designed representations and input-features, to make the best prediction. Automatically learning relevant features and multilevel representation of increasing complexity/abstraction is an emerging field in computer science known as Deep Learning (Grefenstette et al., 2014). Deep learning is mostly based on Deep Belief Networks (DBNs), a technique that consists of having deeper structures of neural networks, now possible due to: more powerful machines, more data, new unsupervised pre-training methods, more efficient methods for parameter estimation, and better parameter regularization techniques. DBN-based approaches are now achieving performances above the state of the art in many areas, and make it now possible to learn how to perform tasks, such as: describe images with sentences (Socher & Lin, 2011; Socher et al., 2014) or associate textual descriptions to entities (Iyyer et al., 2014). The next decade will reveal us how these recent techniques can help us to better process and understand Big Data.

## 3. PERFORMANCE OF PROCESSING AND NETWORKING (CLOUD) INFRASTRUCTURES

This section discusses the huge impact of Big Data in the networks and it has three main parts. Initially, we clearly identify the reasons why Big Data is entailing innovation in the networking arena. In fact, the transmission of large quantities of data poses significant challenges to the current network infrastructures. Additionally, networks that have been around but neither recognized nor well understood, using Big Data analysis, are becoming visible and have been clearly identified as "new networks" (e.g. social networks organized by high-level predicates of keywords). These discovered networks are normally very difficult to manage in a successful way because they have both different features and interrelations among the nodes/agents forming those networks. Then, in this section, we proceed to the second part where are identified some typical scenarios where new network requirements could potentially emerge as a consequence of Big Data exploitation. Finally, we point out some future directions in how to successfully manage the most relevant challenges imposed by Big Data applications on the networking and computing infrastructures.

## Why Big Data is entailing innovation in networks

With the help of Big Data, it is now possible to envisage networks anywhere with strong interrelationships and interactions among them: from the macroscopic to the microscopic worlds; from the global finance to the social human behavior; from species evolution to DNA-level protein interaction; from legacy network infrastructures to others completely different formed by embedded devices, sensors, and very intelligent "self-x" algorithms which can evolve (learn and adjust) as, for example, humans normally do. Nevertheless, these novel network requirements should be managed in much more complex ways than the legacy networks due to some challenging features such as: ubiquitous network access, scalability of network resources (i.e. capacity, delay), abstraction/detail (i.e., what is the more convenient layer to manage the network?), heterogeneity, self-organized network infrastructures, and awareness of the context surrounding each network. Additional novel network requirements are choosing the most suitable set of management policies (e.g. reactive, predictive) to operate the emergent networks with a good performance, and learning new skills from the obtained feedback after a policy has been applied to a network infrastructure, helping to evolve the management policies.

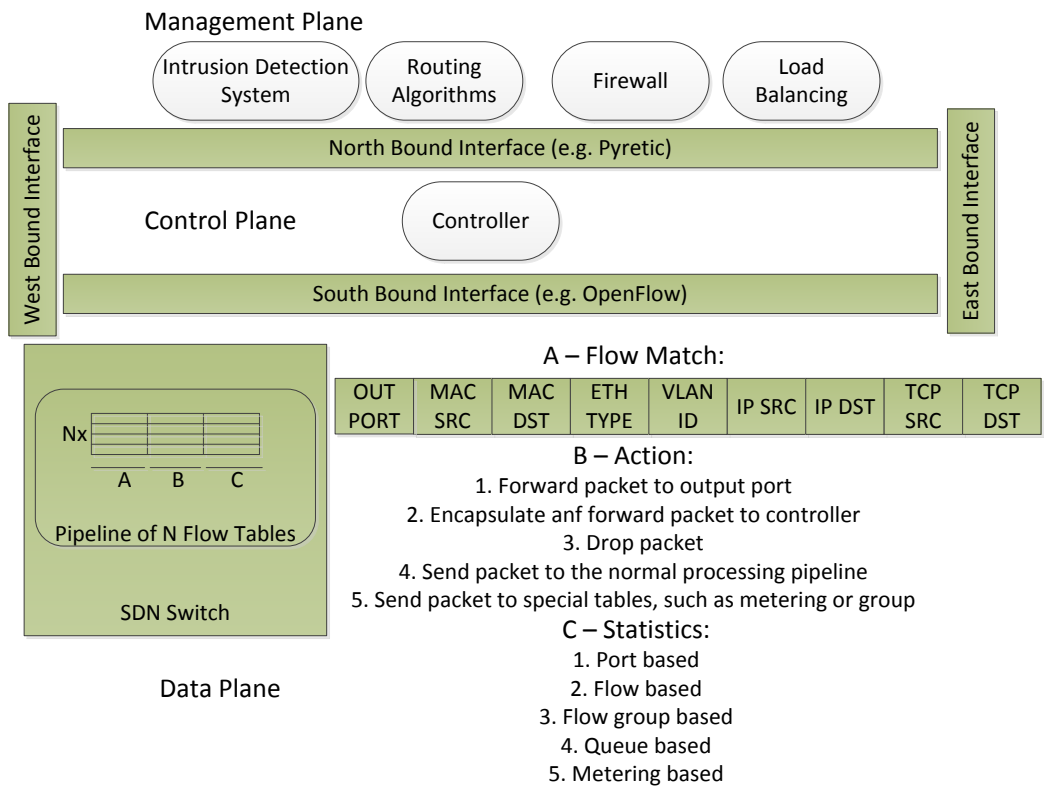

**Figure 2** – OpenFlow-enabled SDN device.

Whichever ways the networks will evolve, the future networks will mainly involve nodes and connecting arcs (Hurlburt & Voas, 2014). Hence, future networks can be discretely measured, traversed, and described using graph theory. Also, a very important aspect to guarantee the success of Big Data is to control with more intelligence and efficiency its computing and networking infrastructures. In parallel, a very recent area to manage networks, which has a huge growth in popularity, is Software Defined Networking (Kreutz & Ramos, 2014; Lara, Kolasani, & Ramamurthy, 2014; Jarraya, 2014; Nunes et al., 2014; Feamster, Rexford, & Zegura, 2013). Using SDN it is possible to decouple the data plane from the control plane, as shown in Figure 2. In this way, the intelligent algorithms to forward/routing traffic are removed from the network devices (e.g. switches, routers) and installed in diverse controllers, potentially in a hierarchical design to satisfy both requirements of scalability and robustness to failures. These controllers can receive from higher levels some management policies via North bound interfaces. These policies are translated by the controllers to low-level pairs of <match, action> that are installed at the network devices through South bound interfaces, using a protocol like either Openflow or Ovsdb, among others. The controllers can also remotely manage the diverse network devices in a coordinated way using West/East bound interfaces. In case of any failure on network nodes or links, the SDN controllers can find alternative data paths traversing the network infrastructure in a more quick and orderly way than legacy networks because the latter rely uniquely in distributed algorithms (which spends more time to converge after a failure to a new network configuration). The next four subsections discusses usage scenarios where SDN can be useful to prepare networks to satisfy novel requirements namely in data centers, Internet interdomain routing, mobile communications, and home networks. A final subsection highlights future SDN developments.

## Data centers

Ultimately data-intensive and collaborative applications are increasingly being moved towards Data Centers aggregated inside the cloud. Here, using Big Data analytics is possible to process data from multiple sources and on-the-fly extract relevant knowledge to drive the strategy or business of either enterprises or other organizations. Consequently to obtain a good performance in the supporting infrastructure for processing big quantities of data, such as low latency and high throughput, some management enhancements are needed to operate more intelligently the available computing resources in each data center, such as virtual machines (Dai et al., 2013), memory (Zhou & Li, 2013), CPU scheduling (Bae et al., 2012), and cache (Koller et al., 2011), I/O (Ram et al., 2013). Other very important functional aspect to be aware in data centers is to enhance the network performance (Marx, 2013; Lange et al., 2011; Saleem, Hassan, & Asirvadam, 2011). To enhance this performance in a more flexible way, the network resources should become virtualized. In this way, we need a "virtualized hypervisor", similar to the one used by virtual machines, to manage the virtual network functions. A SDN controller can support this network hypervisor. That is why the integration of SDN in multi-tenant Data Centers, where there is a significant proliferation of data from all sorts of sources to be processed by Big Data tools (Juniper, 2013), is a very interesting and challenging area to be studied (Koponen et al., 2014; Greenberg et al., 2009; Niranjan et al., 2009). In order to apply SDN to the network infrastructure of data centers, some intelligent software routines should be running in SDN controllers to support the following functionalities: efficient layer 2 switching among virtual and physical switches (IETF, Transparent Interconnection of Lots of Links - TRILL); load balancing; network flow slicing to ensure independence among flows; enforce legacy security functions but now in a completely distributed and dynamic way (i.e. across distinct cloud providers, supplying services to mobile terminals) such as firewalls, Virtual Private Networks (VPNs), DeMilitarized Zones to publicly make services available in a secure way from servers connected within a enterprise domain; authentication; privacy; data integrity; intrusion detection; intrusion prevention; and attack countermeasures. A very recent work by (Perry et al., 2014) proposes to use a similar design to a SDN system but implemented inside the Linux Kernel to provide at a datacenter network several properties such as, low latency, high throughput, fair allocation of network resources between users and applications, deadline-aware scheduling, and congestion (loss) avoidance. The proposed architecture has a centralized arbiter equivalent to a SDN controller. The arbiter basically decides when each packet should be transmitted and what path it should follow. With these two decisions made in the right away for each packet the authors argue that they can satisfy the properties listed above.

## Wide area network (WAN)

A second scenario where SDN can be very useful is the one related with the more suitable control of flow switching in Internet exchange points (IXPs) either among distinct tier-x Internet providers (Feamster, Rexford, & Zegura, 2013) or owned by a single entity with a very large network infrastructure (Jain, Kumar, & Mandal, 2013). An IXP is a location where normally multiple Internet service providers connect theirs networks to exchange routes, often through a common layer-2 switching fabric. The current Internet has globally around 300 IXPs. The commonly used routing protocol to exchange traffic in these IXPs is Border Gateway Protocol (BGP) (Cisco, 2008). However, BGP shows a significant limitation: it was conceived mainly to announce route paths to Internet destination prefixes, and as such it lacks more fine-grained decisions to route wide area traffic. For example, the routing of traffic based on the application type which originated that traffic should have an enormous interest to the efficient transmission of Big Data heterogeneous traffic. In this way, the current BGP needs to be enhanced by high-level policies at the IXP to implement distinct routing policies according the traffic characteristics (e.g. video vs. non-video). SDN exchange point (SDX) (Feamster, Rexford, & Zegura, 2013) could be used for this goal and much more. At the time of this writing, the broadest implementation of SDN in the Internet that is successfully working is described in (Jain, Kumar, & Mandal, 2013). In this work, SDN was used to control the traffic exchange among data centers owned by a single entity with mainly well-specified traffic patterns. So, an interesting question that remains to be answered is if the same SDN solution could be also successfully used in scenarios where the IXPs are traversed by traffic with much more dynamic behavior and from diverse providers. Finally, SDN could be applied at core transport (optical) networks (Azodolmolky et al., 2012) and optical multiplexing to interconnect data centers (Cyan, 2013). For further information, the reader could consult some documentation about version 1.4 of Openflow, which provides support for transactional rule changes, management of switch table space, and supports optical fiber ports (The Open Networking Foundation, 2013; Ren & Xu, 2014).

## Mobile networks

A third scenario where SDN can be very useful is the one related with mobile networks (Costa-Requena, 2014). The major challenge in future mobile networks is how to improve throughput to support the increased demand in data traffic and the decrease demand on the Average Revenue Per User (ARPU), using the current network capabilities in innovative and efficient ways. SDN could help on these issues, redesigning and/or using the current network architectures in novel ways as we discuss in the following text through several relevant examples extracted from the literature.

The first example to enhance mobile networks is to split the network edge from the core (Casado et al., 2012), with the latter forming the fabric that transports packets as defined by an intelligent edge, to software-defined IXPs (Feamster, Rexford, & Zegura, 2013). The second example is to gradually move most of the current LTE network core elements (e.g. middleboxes) to the cloud (Costa-Requena, 2014) in a dynamic way accordingly the load behavior (e.g. data pattern mobility associated to commuters along the day) and traffic characteristics (e.g. security, video transcoding). A third example illustrates how to achieve an efficient access to Big Data infrastructures through mobile devices (Wang, Liu, & Soyata, 2014; Soyata et al., 2014) potentially adjusted to network limitations on the network edge (e.g. terminals, radio access) A fourth example describes how to support an efficient flow forwarding (Pentikousis, Wang, & Hu, 2013) guaranteeing a complete isolation among them in the data path (i.e. using FlowVisor (Sherwood et al., 2010)). A fifth and last example discusses how new techno-economic models can decrease exploitation cost (Naudts et al., 2012) or eventually giving peering (collaboration) incentives to operators to enhance resources usage (e.g. offloading data traffic from a busy access technology to a less-busy one).

## Home networks

A fourth scenario is the one related with home networks (Sundaresan et al., 2011; Kumar, Gharakheili, & Sivaraman, 2013; Yiakoumis et al., 2011). An SDN-based broadband home connection can simplify the addition of new functions in measurement systems such as BISmark (Sundaresan et al., 2011), allowing the system to react to changing conditions in the home network (Kumar, Gharakheili, & Sivaraman, 2013). As an example, a home gateway can perform reactive traffic shaping, considering the current measurement results of the home network. Recent work has proposed slicing control of home networks, to allow different third-party service providers (e.g., smart grid operators) to deploy services on the network without having to install their own infrastructure (Yiakoumis et al., 2011).

## Future directions

To complement the considered usage scenarios some future directions are now discussed regarding the network evolution to successfully satisfy via SDN the challenging requirements raised by Big Data applications, supported by distributed and parallel data processing tools, such as MapReduce (Dean & Ghemawat, 2008), Hadoop (The Apache Software Foundation, 2014a), and visualization tools of large sets of data (GenMAPP (Gladstone Institute University of California at San Francisco, 2014), DIVE (Rysavy, Bromley, & Daggettet, 2014)). In this chapter we advocate that the incorporation of SDN in the networks will be a gradual evolutionary process by two main reasons. The first one is related to the existence of diverse network players. As an example, in the current Internet, there are at least three major ones, the Content Network providers, the core Autonomous Systems (ASs), and the edge operators. In this way, it seems natural that each one of these three entities could adopt SDN at different times and using diverse deployment strategies, depending on each one particular needs and already deployed solutions. This vision is also discussed in (Vissicchio, Vanbever, & Bonaventure, 2014). They discuss a gradual global strategy adoption of SDN based on which network layer and whose entity is responsible to control via SDN the Forwarding Information Base (FIB), i.e. forwarding tables, at distinct node locations of the network infrastructure. Obviously the final step of this evolutionary process is the full integration of SDN among all the existing entities. The second reason why SDN deployment should be a gradual process is concerned with system reliability after the adoption of a new SDN solution. By this, we mean that a new adopted SDN solution operating in a real network infrastructure should operate like it was initially planned for and it should not degrade the network performance when compared with previous solutions. To avoid this, each new SDN solution can be currently troubleshooted and verified in each one of the following distinct SDN aspects: static analysis verification of network configuration – Router Configuration Checker (Feamster & Balakrishnan, 2005) to avoid routing partitions and loops, static data plane verification by performing symbolic execution on packets to check flow isolation - Header Space Analysis (Kazemian, Change, & Zheng, 2013), and dynamic verification of control plane sequential composition of Finite State Machines triggered by JSON events – Kinetic (Monsanto et al., 2013). Nevertheless, after all these verifications have been successfully made, we are not always completely sure that after the deployment of a new SDN solution the network will be working completely out of problems due to two important limitations such as: each verification (e.g. data plane vs. control plane) has been made in a completely isolated way from the others; some unexpected interactions could arise from the dynamic interaction among several controllers with distinct purposes when they are in operation. In this way, an interesting area of research is to study how to perform a single-step robust troubleshooting and verification of a complete (i.e., configuration, data plane, control plane) SDN system in a similar way as proposed in (Nelson et al., 2014), including the verification of fully dynamic interactions among controllers with potential distinct goals (e.g., load balancing, security, forwarding) and fulfilling very complex policies formed by the composition (i.e., applying diverse operators such as parallel, sequential, conditional if) of other policies at high-level layers of abstraction (Foster et al., 2013).

Finally, another open issue for research is how to collect in real-time a huge amount of forensic data from the network operation, through, for example, packet inspection, traffic captures, log files, and report files, to find the root causes of different problems. All this may enable a more efficient and intelligent way to control the global network, such as the solutions given by SDN. Using this work direction, we assume the network infrastructure as a valuable producer of Big Data information to be analyzed and explored in unpredictable directions.

## 4. A CASE STUDY FOR AN INTELLIGENT INTERNET EXCHANGE POINT

This section discusses and evaluates an intelligent solution to manage an Internet Exchange Point (IXP). An IXP is a location on the network where normally different operator networks (also designated by Autonomous Systems – ASs) exchange traffic with each other. As discussed before there is a growing interest in applying SDN to make some aspects of wide-area network management easier and more flexible by giving operators direct control over packet-processing rules that match on multiple header fields and perform a variety of potentially very interesting management actions.

In this section, we evaluate an implementation of a software-defined IXP (SDX) (Feamster, Rexford, & Zegura, 2013). This implementation provides new programming abstractions allowing BGP (Cisco, 2008) participants to create and rule new wide area traffic delivery applications (e.g., differentiate video from other traffic), avoiding distinct applications to interfere with each other and routing their traffic with distinct high-level management policies (Feamster, 2014). This implementation acts like a controller that uses policies specified in Pyretic (Contributors, 2013b).

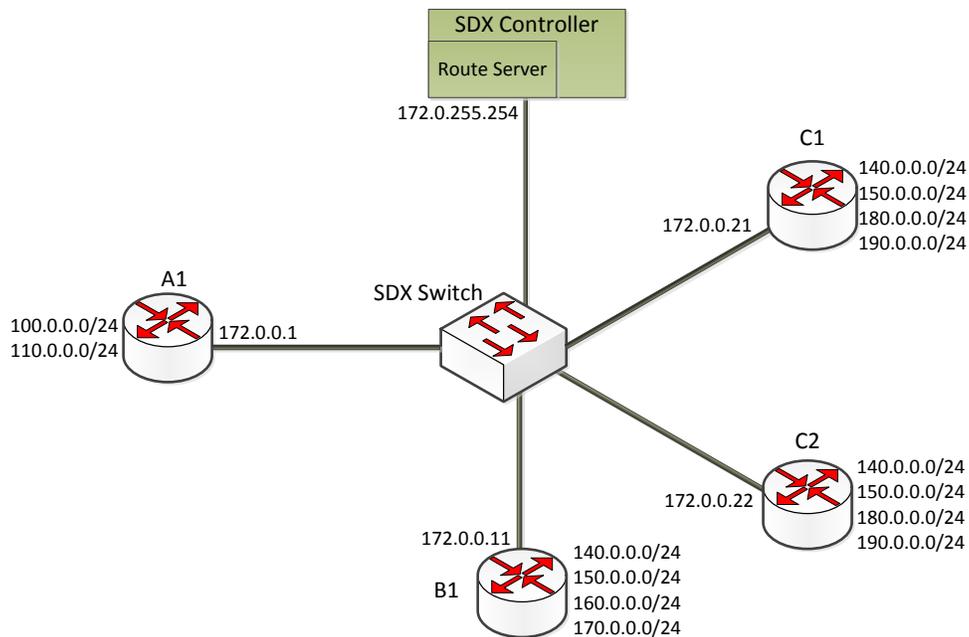

**Figure 3** – BGP scenario (Feamster, 2014).

In this case study, we aim to implement an intelligent and efficient IXP controller module, which should have the following behavior in an IXP-BGP scenario (see Figure 3) with three peering ASs (i.e., A, B, and C):

- Each AS establishes a single BGP/TCP connection with only the route server of the SDX controller and not with any other neighbor AS; this is to guarantee that the route server is the only entity in the network that can populate the routing tables of the AS´s edge routers but doing so the routing rules should satisfy both pyretic policies specified in the controller and the announced networks (to the route server) by each AS;
- AS A's policy forwards web traffic (destined to TCP port 80) to peer B;

- AS A´s policy forwards traffic destined to ports 443 (HTTPS) and 8080 to peer C;
- AS B has no defined pyretic policy, so its forwarding proceeds to default BGP forwarding;
- AS C has two input ports at the SDX. Consequently, the controller has a first policy that forwards HTTPS traffic to the AS C´s input port [0] for c1, a second policy that forwards traffic destined to port 80 to the AS C´s input port [1] for c2, and a third policy that drops any traffic for port 8080;
- When the route server associated to the SDX controller learns more than one BGP route to the same destination prefix, the route server decides the best path that was announced by a BGP participant applying a pre-defined criterion (e.g., choose the one with the smallest router-id).

## Understanding the SDX setup

The network topology visualized in Figure 3 will be evaluated using the network emulator mininext (Schlinker, 2014), Quagga router (Contributors, 2013a) to deploy routing with BGP (Cisco, 2008), and the SDX framework (Feamster, Rexford, & Zegura, 2013) to implement the already discussed management policies on top of BGP. Table 1 shows a partial view of the mininext setup file where each host interface is configured. Each host is configured in such a way it is within a specific AS domain.

**Table 1 -** Configuring host interface in mininext setup file.

```
print "Configuring host interface; each host belongs to a specific AS domain"
    for host in hosts:
        print "Host name: ", host.name
        if host.name=='a1':
           host.cmd('sudo ifconfig lo:1 100.0.0.1 netmask 255.255.255.0 up')
                       …
        if host.name=='b1':
           host.cmd('sudo ifconfig lo:140 140.0.0.1 netmask 255.255.255.0 up')
                       …
```

Table 2 illustrates the BGP routing configuration for participant A (file bgpd.conf). We briefly explain some configuration details: the BGP id of AS A is 100; the A edge router-id is 172.0.0.1; the BGP´s only neighbor is the SDX router, so the networks within A´s domain are only announced to that router; the networks belonging to A´s domain are 100.0.0.0/24 and 110.0.0.0/24.

**Table 2 -** BGP routing configuration for participant A.

```
router bgp 100
 bgp router-id 172.0.0.1
 neighbor 172.0.255.254 remote-as 65000
 network 100.0.0.0/24
 network 110.0.0.0/24
 redistribute static
```

The SDX presents to each participant (A, B, C) a virtual SDX switch. Each participant specifies pyretic policies for its virtual switch independently of other participants' policies. This limited view of the network ensures that the participants are not allowed to write rules for other network's traffic. For more details consult (Feamster, Rexford, & Zegura, 2013). Table 3 shows the outbound SDX management policies specified (in pyretic) for participant A. These policies were already discussed. By outbound policy we mean a policy to be applied to traffic entering a virtual switch on a physical port from the participant's own border router.

**Table 3 -** Outbound SDX management policies specified (in pyretic) for participant A.

```
final_policy = ((match(dstport=80)  >> sdx.fwd(participant.peers['B'])) +
                (match(dstport=443) >> sdx.fwd(participant.peers['C'])) +
                (match(dstport=8080)>> sdx.fwd(participant.peers['C'])))
```

Table 4 shows inbound SDX management policies specified (in pyretic) for participant C. These policies were already discussed. By inbound policy we mean a policy to be applied to the traffic entering a virtual switch on a virtual port from another SDX participant. Note also that it is not necessary to specify explicitly a drop policy to traffic destined to port 8080 because the default SDX policy is drop a packet if that packet does not have a positive match with any present pyretic policy.

**Table 4 -** Inbound SDX management policies specified (in pyretic) for participant C.

```
final_policy = ((match(dstport=443) >> sdx.fwd(participant.phys_ports[0])) +
                (match(dstport=80) >> sdx.fwd(participant.phys_ports[1])))
```

## Evaluation

The network topology used in the current evaluation made with a network emulator is shown in Figure 3. All the evaluation was performed in a single Linux virtual machine (Ubuntu).

After the configuration process has been finalized, we can now initiate the evaluation. For that, one should open a Linux shell, and run the pyretic SDX controller with the following command:

```
./sdx-setup.sh pyretic
```

In a separate console, launch the Mininext (Schlinker, 2014) topology as follows:

```
./sdx-setup.sh mininet app_specific_peering_inboundTE
```

In a separate console, launch the SDX route server as follows:

```
./sdx-setup.sh exabgp
```

At this step one can check whether the participants received the routes from the SDX route server. For example, to verify the available routes on host a1, in the console running mininext, one can perform the following command:

```
mininext> a1 route -n
```

The output of previous command should be similar to the one visualized in Figure 4.

```
/bgpd --daemon -A 127.0.0.1
root        36  0.0  0.2  18692  1312 pts/7    R+   07:51   0:00 ps aux
**Adding Network Interfaces for SDX Setup
Configuring participating ASs

Host name:  a1
Host name:  b1
Host name:  c1
Host name:  c2
Host name:  exabgp
** Running CLI
*** Starting CLI:
mininext> a1 route -n
Kernel IP routing table
Destination     Gateway         Genmask         Flags Metric Ref    Use Iface
140.0.0.0       172.0.1.3       255.255.255.0   UG    0      0        0 a1-eth0
150.0.0.0       172.0.1.3       255.255.255.0   UG    0      0        0 a1-eth0
160.0.0.0       172.0.1.2       255.255.255.0   UG    0      0        0 a1-eth0
170.0.0.0       172.0.1.2       255.255.255.0   UG    0      0        0 a1-eth0
172.0.0.0       0.0.0.0         255.255.0.0     U     0      0        0 a1-eth0
180.0.0.0       172.0.1.4       255.255.255.0   UG    0      0        0 a1-eth0
190.0.0.0       172.0.1.4       255.255.255.0   UG    0      0        0 a1-eth0
mininext>
```

**Figure 4** – Routing table for node a1 (edge router of AS A).

At this point, it is necessary to clarify the ip addresses that appear as the next-hop (gateway) address in each one of the routing table lines of Figure 4. To avoid flow table become too large for the SDX's switch, the SDX controller introduces the concept of Virtual Next Hops (VNHs). SDX platform assigns one (virtual) next hop for each set of IP prefixes with similar forwarding behavior. For example, in this example IP prefix pair (160.0.0.0/24, 170.0.0.0/24) have similar forwarding behavior. Thus the controller assigns a single VNH for that pair. You can verify this behavior from the output messages from Pyretic's console shown in Table 5, from where we have selected the following information:
Virtual Next Hop --> IP Prefix: {'VNH2': set([u'160.0.0.0/24', u'170.0.0.0/24']), …
Virtual Next Hop --> Next Hop IP Address (Virtual): {'VNH2': '172.0.1.2', ...
The previous selected information shows that the SDX controller assigns (160.0.0.0/24, 170.0.0.0/24) to VNH2, which has the next-hop address such as 172.0.1.2. Refer to (Feamster, Rexford, & Zegura, 2013) for more details on Virtual Next Hops.

**Table 5 -** Output messages from Pyretic's console related with VNH configuration

```
Parsing participant's policies
Starting VNH Assignment
After new assignment
Virtual Next Hop --> IP Prefix:
 {'VNH1': set([u'110.0.0.0/24', u'100.0.0.0/24']),
  'VNH2': set(['160.0.0.0/24', '170.0.0.0/24']),
  'VNH3': set(['140.0.0.0/24', '150.0.0.0/24']),
  'VNH4': set(['190.0.0.0/24', '180.0.0.0/24'])}
Virtual Next Hop --> Next Hop IP Address (Virtual):
 {'VNH1': '172.0.1.1', 'VNH2 ': '172.0.1.2', 'VNH3': '172.0.1.3', 'VNH4': '172.0.1.4',
 'VNH': [IPAddress('172 .0.1.0'), IPAddress('172.0.1.1'), IPAddress('172.0.1.2'),
    IPAddress('172.0.1.3'), IPAddress('172.0.1.4'), IPAddress('172.0.1.5'),
    IPAddress('172.0.1.6'), IPAddress('172.0.1.7'), IPAddress('172.0.1.8'),
    IPAddress('172.0.1.9'), IPAddress('172 .0.1.10'), IPAddress('172.0.1.11'),
    IPAddress('172.0.1.12'), IPAddress('172.0.1. 13'), IPAddress('172.0.1.14'),
    IPAddress('172.0.1.15')]}
Virtual Next Hop --> Next Hop Mac Address (Virtual)
 {'VNH1': aa:00:00:00:00:01, 'VNH2': aa:00:00:00:00:02, 'VNH3': aa:00:00:00:00:03,
  'VNH4': aa:00:00:00:00:04, 'VNH': 'aa:00:00:00:00:00'}
Completed VNH Assignment
```

The next evaluation step is to check if the pyretic policies specified before can be correctly applied in our system. For this, the iperf tool can be used in the mininext console. The diverse issued commands and corresponding results are visualized and discussed as follows. From Figure 5 one can conclude that port 80 traffic for routes advertised by AS B is received by node b1.

```
/bgpd --daemon -A 127.0.0.1
root       33  0.0  0.2 18692 1312 pts/7    R+   11:28   0:00 ps aux
**Adding Network Interfaces for SDX Setup
Configuring participating ASs

Host name:  a1
Host name:  b1
Host name:  c1
Host name:  c2
Host name:  exabgp
** Running CLI
*** Starting CLI:
mininext> b1 iperf -s -B 140.0.0.1 -p 80 &
mininext> a1 iperf -c 140.0.0.1 -B 100.0.0.1 -p 80 -t 2
------------------------------------------------------
Client connection to 140.0.0.1, TCP port 80
Binding to local address 100.0.0.1
TCP window size: 85.3 KByte (default)
------------------------------------------------------
[ 3] local 100.0.0.1 port 80 connected with 140.0.0.1 port 80
[ ID] Interval       Transfer     Bandwidth
[ 3]  0.0- 2.0 sec  78.5 MBytes   329 Mbits/sec
mininext>
```

**Figure 5** – Test routing path to AS B (server destination port is 80).

From Figure 6, one can verify that port 80 traffic from AS A for routes advertised only by AS C is forwarded to node c2.

```
Host name:  exabgp
** Running CLI
*** Starting CLI:
mininext> b1 iperf -s -B 140.0.0.1 -p 80 &
mininext> a1 iperf -c 140.0.0.1 -B 100.0.0.1 -p 80 -t 2
------------------------------------------------------
Client connecting to 140.0.0.1, TCP port 80
Binding to local address 100.0.0.1
TCP window size: 85.3 KByte (default)
------------------------------------------------------
[  3] local 100.0.0.1 port 80 connected with 140.0.0.1 port 80
[ ID] Interval       Transfer     Bandwidth
[  3]  0.0- 2.0 sec  78.5 MBytes   329 Mbits/sec
mininext> c2 iperf -s -B 180.0.0.1 -p 80 &
mininext> a1 iperf -c 180.0.0.1 -B 100.0.0.2 -p 80 -t 2
------------------------------------------------------
Client connecting to 180.0.0.1, TCP port 80
Binding to local address 100.0.0.2
TCP window size: 85.3 KByte (default)
------------------------------------------------------
[  3] local 100.0.0.2 port 80 connected with 180.0.0.1 port 80
[ ID] Interval       Transfer     Bandwidth
[  3]  0.0- 2.0 sec  78.2 MBytes   327 Mbits/sec
mininext>
```

**Figure 6** – Test routing path to AS C (server destination port is 80).

Finally, from Figure 7, one can also verify that the port 8080 traffic forwarded to AS C is dropped as the last client iperf message elucidates: "Connection time out". In this way, we finalize our case study.

The current case study clearly illustrates that using high-level and intuitive management policies is possible to easily route data traffic with diverse requirements (e.g. video vs. non-video) among network infrastructures owned by several entities, which is a very realistic scenario to deploy Big Data heterogeneous services in the wide area network through diverse public cloud providers. In this way, it is possible to optimize the global network performance, reduce the network costs and proactively managing customer experience.

```
mininext> b1 iperf -s -B 140.0.0.1 -p 80 &
mininext> a1 iperf -c 140.0.0.1 -B 100.0.0.1 -p 80 -t 2
------------------------------------------------------
Client connecting to 140.0.0.1, TCP port 80
Binding to local address 100.0.0.1
TCP window size: 85.3 KByte (default)
------------------------------------------------------
[  3] local 100.0.0.1 port 80 connected with 140.0.0.1 port 80
[ ID] Interval       Transfer     Bandwidth
[  3]  0.0- 2.0 sec  78.5 MBytes   329 Mbits/sec
mininext> c2 iperf -s -B 180.0.0.1 -p 80 &
mininext> a1 iperf -c 180.0.0.1 -B 100.0.0.2 -p 80 -t 2
------------------------------------------------------
Client connecting to 180.0.0.1, TCP port 80
Binding to local address 100.0.0.2
TCP window size: 85.3 KByte (default)
------------------------------------------------------
[  3] local 100.0.0.2 port 80 connected with 180.0.0.1 port 80
[ ID] Interval       Transfer     Bandwidth
[  3]  0.0- 2.0 sec  78.2 MBytes   327 Mbits/sec
mininext> c1 iperf -s -B 180.0.0.1 -p 8080 &
mininext> a1 iperf -c 180.0.0.1 -B 100.0.0.1 -p 8080 -t 2
connect failed: Connection timed out
mininext>
```

**Figure 7** – Test routing path to AS C (server destination port is 8080).

## 5. CONCLUSION

This chapter discusses two important aspects related to a successful deployment of a Big Data infrastructure. The first aspect concerns the extraction of relevant knowledge from heterogeneous and massive amounts of data. The second aspect relates to new challenges that Big Data applications and services are imposing in network infrastructures.

Machine Learning methods and related technologies can be used to extract relevant knowledge from the large and diverse datasets currently available. From a Machine Learning point of view, we have seen that usually more data means more reliable models and results. The chapter focused on the massive amounts of unstructured textual information that are constantly being produced, and that can be accessed from online sources, which constitute a potential valuable source of knowledge for companies, organizations, and governments. Text Mining is described as a recent research area that is responsible for extracting useful information from such unstructured texts, drawing knowledge and techniques from different fields of computer science. Today's text mining research and technology enables high-performance analytics from web's textual data, allowing to: cluster documents and web pages according to their content, find associations among entities (people, places and/or organizations), and reasoning about important data trends.

The challenges being imposed in the network infrastructure to efficiently transmit Big Data, which is typically heterogeneous and distributed, were also discussed as follows: decide about the best deployment strategy for new network management solutions; choose the most correct method to fully troubleshoot and verify, in an automatic way, each new network management solution before its real implementation; and collect real-time forensic data from the network operation to supervise and control the network infrastructure in both novel and efficient ways.

Finally, a practical scenario concerning the efficient transmission of Big Data heterogeneous traffic through the wide area network among diverse public cloud providers was presented. In this way, we have studied a solution that enhances the current Internet inter-AS routing protocol (i.e. BGP) with high-level management policies at the IXP to implement distinct routing policies according the traffic characteristics (e.g. video vs. non-video). This solution has been also validated and, from the obtained results, we can conclude that the default BGP routing, only based in the destination IP prefixes, can be complemented with further information (e.g. destination port). This last characteristic enables traffic routing also based in application details, which can enhance the customer experienced quality.

## KEY TERMS AND DEFINITIONS

Big Data: the term that represents data sets that are extremely large to handle through traditional methods. Big data represents information that has such a high volume, velocity, variety, variability, veracity and complexity that require specific mechanisms to produce real value from it in a timely way.

Intelligent Data Management: a set of solutions to help organizations reduce cost, complexity, and time when they aim to analyze their data in order to extract some well identified usefulness.

Text Mining: a process of extracting relevant knowledge from large collections of unstructured text documents. In this way, text mining usually involves the process of structuring the input text, deriving patterns within the structured data, and finally evaluation and interpretation of the output.

Machine Learning: it is a type of Artificial Intelligence (AI) that provides computers with the ability to learn without being explicitly programmed. Machine learning explores the construction and study of algorithms that can learn from and make predictions on data.

IXP: it is an Internet location where normally multiple Internet service providers connect theirs networks to exchange traffic messages. This exchange is made possible by a routing path vector protocol, i.e. BGP.

JSON: the Javascript Object Notation (JSON) is a language-independent and open data format that can be used to transmit human-readable text-based object information, across domains, using an attribute-value pair's notation and easy-to-access manner.

SDN: Software Defined Networking (SDN) allows logically centralized controllers to manage network services through the decoupling of system control from the underlying traffic exchange. Some advantages of using SDN are decreasing the maintenance cost and fostering innovation on the networking infrastructures.

**REFERENCES**

Asuncion, A., & Newman, D. J. (2007). UCI Machine Learning Repository. *University of California Irvine School of Information*. Retrieved May 15, 2015, from http://www.ics.uci.edu/~mlearn/MLRepository.html

Azodolmolky, S., Nejabati, R., Peng, S., Hammad, A., Channegowda, M. P., Efstathiou, N., … Simeonidou, D. (2012). Optical FlowVisor: An OpenFlow-based Optical Network Virtualization Approach. In *National Fiber Optic Engineers Conference* (p. JTh2A.41). Optical Society of America. Retrieved May 15, 2015, from http://www.opticsinfobase.org/abstract.cfm?URI=NFOEC-2012-JTh2A.41

Bae, C., Xia, L., Dinda, P., & Lange, J. (2012). Dynamic Adaptive Virtual Core Mapping to Improve Power, Energy, and Performance in Multi-socket Multicores. In *Proceedings of the 21st International Symposium on Applied Computing* (pp. 247–258).

Bollen, J., Mao, H., & Zeng, X. (2011). Twitter mood predicts the stock market. *Journal of Computational Science*, *2*(1), 1–8.

Casado, M., Koponen, T., Shenker, S., & Tootoonchian, A. (2012). Fabric: a retrospective on evolving SDN. *Hot Topics in Software Defined Networking (HotSDN)*, 85–89. Retrieved May 15, 2015, from http://dl.acm.org/citation.cfm?id=2342459

Cisco. (2008). Border Gateway Protocol. *Update*. Retrieved April 2, 2014, from http://docwiki.cisco.com/wiki/Border_Gateway_Protocol

Contributors. (2013a). Quagga Routing Suite. *Online*. Retrieved May 15, 2015, from http://www.nongnu.org/quagga/

Contributors. (2013b). The Policy: Pyretic's Foundation. Retrieved April 18, 2014, from https://github.com/frenetic-lang/pyretic/wiki/Language-Basics

Costa-Requena, J. (2014). SDN integration in LTE mobile backhaul networks. In *International Conference on Information Networking* (pp. 264–269).

Cox, M., & Ellsworth, D. (1997). Application-controlled demand paging for out-of-core visualization. *Proceedings. Visualization '97 (Cat. No. 97CB36155)*.

Cyan. (2015). Z-Series. *Cyan*. Retrieved May 13, 2015, from http://www.cyaninc.com/products/z-series-packet-optical

Dai, Y., Qi, Y., Ren, J., Shi, Y., Wang, X., & Yu, X. (2013). A lightweight VMM on many core for high performance computing. *ACM SIGPLAN Notices*, *48*(7), 111-120. Retrieved May 15, 2015, from http://dl.acm.org/citation.cfm?doid=2517326.2451535

Davenport, T. H., Harris, J. G., & Morison, R. (2010). *Analytics at Work: Smarter Decisions, Better Results*. *Harvard Business School Press Books*. Retrieved May 15, 2015, from http://www.amazon.com/dp/1422177696

Davenport, T. H., & Patil, D. J. (2012). Data scientist: the sexiest job of the 21st century. *Harvard Business Review*, *90*(10), 70-76.

Dean, J., & Ghemawat, S. (2008). MapReduce : Simplified Data Processing on Large Clusters. *Communications of the ACM*, *51*(1), 1–13. doi:10.1145/1327452.1327492

Diebold, F. X. (2012). A Personal Perspective on the Origin(s) and Development of "Big Data": The Phenomenon, the Term, and the Discipline, Second Version. *SSRN Electronic Journal*, (13-003). Retrieved May 15, 2015, from http://econpapers.repec.org/RePEc:pen:papers:13-003

Feamster, N. (2014). Software Defined Networking. Retrieved July 15, 2014, from Available from https://www.coursera.org/course/sdn

Feamster, N., & Balakrishnan, H. (2005). Detecting BGP configuration faults with static analysis. *Proc. Networked Systems Design and Implementation*, 49–56. Retrieved May 15, 2015, from http://www.usenix.org/event/nsdi05/tech/feamster/feamster_html/

Feamster, N., Rexford, J., Shenker, S., Clark, R., Hutchins, R., Levin, D., & Bailey, J. (1986). SDX: A Software-Defined Internet Exchange. *Proceedings IETF 86*. Retrieved May 13, 2015, from http://www.ietf.org/proceedings/86/slides/slides-86-sdnrg-6

Feamster, N., Rexford, J., & Zegura, E. (2013). The Road to SDN. *Queue*, *11*(12), 20–40. Retrieved May 15, 2015, from http://dl.acm.org/citation.cfm?doid=2559899.2560327

Foster, N., Guha, A., Reitblatt, M., Story, A., Freedman, M. J., Katta, N. P., … Harrison, R. (2013). Languages for software-defined networks. *IEEE Communications Magazine*, *51*(2), 128–134.

Gladstone Institute University of California at San Francisco. (2014). GenMAPP. *GenMAPP*. Retrieved May 13, 2015, from http://www.genmapp.org/about.html

Greenberg, B. A., Hamilton, J. R., Kandula, S., Kim, C., Lahiri, P., Maltz, A., … Maltz, D. A. (2009). VL2: a scalable and flexible data center network. In *Proceedings of the ACM SIGCOMM 2009 conference on Data communication* (Vol. 09, pp. 51–62). ACM. Retrieved May 15, 2015, from http://doi.acm.org/10.1145/1592568.1592576

Grefenstette, E., Blunsom, P., Freitas, N. De, & Hermann, K. M. (2014). A Deep Architecture for Semantic Parsing. In *ACL Workshop on Semantic Parsing* (pp. 22–27).

Hall, M., National, H., Frank, E., Holmes, G., Pfahringer, B., Reutemann, P., & Witten, I. H. (2009). The WEKA Data Mining Software : An Update. *SIGKDD Explorations*, *11*(1), 10–18.

Hirschberg, J. (1998). "Every time I fire a linguist, my performance goes up", and other myths of the statistical natural language processing revolution. In *15th National Conference on Artificial Intelligence, Madison, Wisconsin (Invited Speech)*.

Hurlburt, G. F., & Voas, J. (2014). Big data, networked worlds. *Computer*, *47*(4), 84–87.

Iyyer, M., Boyd-Graber, J., Claudino, L., Socher, R., & Daumé III, H. (2014). A Multi-Sentence Neural Network for Connecting Textual Descriptions to Entities. In *Conference on Empirical Methods in Natural Language Processing (EMNLP 2014)*.

Jain, S., Kumar, A., & Mandal, S. (2013). B4: Experience with a globally-deployed software defined WAN. *Sigcomm*, 3–14. Retrieved May 15, 2015, from http://dl.acm.org/citation.cfm?id=2486019

Jarraya, Y. (2014). A Survey and a Layered Taxonomy of Software-Defined Networking. *Communications Surveys & Tutorials, IEEE*, *16*(4), 1955 – 1980. doi:10.1109/COMST.2014.2320094

Jelinek, F. (2005). Some of my best friends are linguists. *Language Resources and Evaluation*, *39*(1), 25–34.

Juniper. (2013). Integrating SDN into the Data Center. *White Paper*. Retrieved May 15, 2015, from http://www.juniper.net/us/en/local/pdf/whitepapers/2000542-en.pdf

Kazemian, P., Change, M., & Zheng, H. (2013). Real Time Network Policy Checking Using Header Space Analysis. *USENIX Symposium on Networked Systems Design and Implementation*, 1–13. Retrieved May 15, 2015, from http://yuba.stanford.edu/~peyman/docs/net_plumber-nsdi13.pdf

Koller, R., Verma, A., & Rangaswami, R. (2011). Estimating application cache requirement for provisioning caches in virtualized systems. In *IEEE International Workshop on Modeling, Analysis, and Simulation of Computer and Telecommunication Systems - Proceedings* (pp. 55–62).

Koponen, T., Amidon, K., Balland, P., Casado, M., Chanda, A., Fulton, B., … Zhang, R. (2014). Network Virtualization in Multi-tenant Datacenters. In *Proceedings of the 11th USENIX Symposium on Networked Systems Design and Implementation (NSDI 14)* (pp. 203–216). USENIX. Retrieved May 15, 2015, from http://blogs.usenix.org/conference/nsdi14/technical-sessions/presentation/koponen

Kreutz, D., & Ramos, F. (2014). Software-Defined Networking: A Comprehensive Survey. *arXiv Preprint arXiv: …*, 49. Retrieved May 15, 2015, from http://arxiv.org/abs/1406.0440

Kumar, H., Gharakheili, H. H., & Sivaraman, V. (2013). User control of quality of experience in home networks using SDN. In *Advanced Networks and Telecommuncations Systems (ANTS), 2013 IEEE International Conference on* (pp. 1–6).

Lange, J. R., Pedretti, K., Dinda, P., Bridges, P. G., Bae, C., Soltero, P., & Merritt, A. (2011). Minimal-overhead virtualization of a large scale supercomputer. *ACM SIGPLAN Notices*.

Lara, A., Kolasani, A., & Ramamurthy, B. (2013). Network Innovation using OpenFlow: A Survey. *IEEE Communications Surveys & Tutorials*, *PP*(99), 1–20. Retrieved May 15, 2015, from http://ieeexplore.ieee.org/lpdocs/epic03/wrapper.htm?arnumber=6587999

Linden, G., Smith, B., & York, J. (2003). Amazon.com recommendations: Item-to-item collaborative filtering. *IEEE Internet Computing*, *7*(1), 76–80.

Marx, V. (2013). Biology: The big challenges of big data. *Nature*, *498*(7453), 255–260. Retrieved May 15, 2015, from http://www.nature.com.gate1.inist.fr/nature/journal/v498/n7453/full/498255a.html\nhttp://www.nature.com.gate1.inist.fr/nature/journal/v498/n7453/pdf/498255a.pdf

McAfee, A., & Brynjolfsson, E. (2012). Big Data. The management revolution. *Harvard Buiness Review*, *90*(10), 61–68. Retrieved May 15, 2015, from http://www.buyukverienstitusu.com/s/1870/i/Big_Data_2.pdf

Michael, K., & Miller, K.W. (2013). Big Data: New Opportunities and New Challenges. Computer, 46(6), 22-24.

Miller, K. W., & St, M. (2013). Big Data : New Opportunities and New Challenges. *Computer*, *46*(6), 22–24. doi:10.1109/MC.2013.196

Monsanto, C., Reich, J., Foster, N., Rexford, J., & Walker, D. (2013). Composing software-defined networks. *Proceedings of the 10th USENIX Conference on Networked Systems Design and Implementation*, 1–14. Retrieved May 15, 2015, from http://dl.acm.org/citation.cfm?id=2482626.2482629\nhttp://www.frenetic-lang.org/pyretic/

Nature. (2008). Community cleverness required. *Nature*, *455*(7209), 1.

Naudts, B., Kind, M., Westphal, F. J., Verbrugge, S., Colle, D., & Pickavet, M. (2012). Techno-economic analysis of software defined networking as architecture for the virtualization of a mobile network. In *Proceedings - European Workshop on Software Defined Networks, EWSDN 2012* (pp. 67–72).

Nelson, T., Ferguson, A. D., Scheer, M. J. G., & Krishnamurthi, S. (2014). Tierless Programming and Reasoning for Software-Defined Networks. In *Proceedings of the 11th USENIX Symposium on Networked Systems Design and Implementation (NSDI 14)* (pp. 519–531). USENIX. Retrieved May 15, 2015, from http://blogs.usenix.org/conference/nsdi14/technical-sessions/presentation/nelson

Niranjan Mysore, R., Pamboris, A., Farrington, N., Huang, N., Miri, P., Radhakrishnan, S., … Mysore, R. N. (2009). PortLand: a scalable fault-tolerant layer 2 data center network fabric. In *SIGCOMM '09 Proceedings of the ACM SIGCOMM 2009 conference on Data communication* (pp. 39–50). ACM. Retrieved May 15, 2015, from http://doi.acm.org/10.1145/1592568.1592575

Nunes, B. A. A., Mendonca, M., Nguyen, X. N., Obraczka, K., & Turletti, T. (2014). A Survey of Software-Defined Networking: Past, Present, and Future of Programmable Networks. *IEEE Communications Surveys and Tutorials*.

Parise, S., Iyer, B., & Vesset, D. (2012). Four Strategies to Capture and Create Value from Big Data. *Ivey Business Journal*, *76*(4), 1–5. Retrieved May 15, 2015, from http://search.ebscohost.com/login.aspx?direct=true&db=bth&AN=78946504&site=bsi-live

Pentikousis, K., Wang, Y., & Hu, W. (2013). Mobileflow: Toward software-defined mobile networks. *IEEE Communications Magazine*, *51*(7), 44–53.

Perry, J., Ousterhout, A., Balakrishnan, H., Shah, D., & Fugal, H. (2014). Fastpass: A Centralized "Zero-Queue" Datacenter Network. In *Sigcomm* (pp. 307-318).

Press, G. (2013). A Very Short History Of Big Data. *Forbes*. Retrieved April 2, 2014, from http://www.forbes.com/sites/gilpress/2013/05/09/a-very-short-history-of-big-data/

Ram, K., Cox, A., Chadha, M., & Rixner, S. (2013). Hyper-Switch: A Scalable Software Virtual Switching Architecture. In *USENIX ATC 2013* (pp. 13–24).

Ren, T., & Xu, Y. (2014). Analysis of the New Features of OpenFlow 1.4. In *2nd International Conference on Information, Electronics and Computer* (pp. 73-77).

Rysavy, S. J., Bromley, D., & Daggett, V. (2014). DIVE: A graph-based visual-analytics framework for big data. *IEEE Computer Graphics and Applications*, *34*(2), 26–37.

Saleem, H. M., Hassan, M. F., & Asirvadam, V. S. (2011). P2P service discovery in clouds with message level intelligence. In *2011 National Postgraduate Conference - Energy and Sustainability: Exploring the Innovative Minds, NPC 2011*.

Schlinker, B. (2014). Mininext. Retrieved May 13, 2015, from https://github.com/USC-NSL/miniNeXT

Sherwood, R., Gibb, G., Yap, K. K.-K. K., Appenzeller, G., Casado, M., McKeown, N., & Parulkar, G. M. (2010). Can the Production Network Be the Testbed? In R. H. Arpaci-Dusseau & B. Chen (Eds.), *9th {USENIX} Symposium on Operating Systems Design and Implementation, {OSDI} 2010, October 4-6, 2010, Vancouver, {BC}, Canada, Proceedings* (Vol. M, pp. 365–378). USENIX Association. Retrieved May 15, 2015, from http://www.usenix.org/events/osdi/tech/full_papers/Sherwood.pdf\nhttp://static.usenix.org/legacy/events/osdi10/tech/full_papers/Sherwood.pdf\nhttp://www.usenix.org/event/osdi10/tech/full_papers/osdi10_proceedings.pdf

Socher, R., Karpathy, A., Le, Q. V, Manning, C. D., & Ng, A. Y. (2014). Grounded Compositional Semantics for Finding and Describing Images with Sentences. *Transactions of the Association for*

*Computational Linguistics (TACL)*, *2*(April), 207–218. Retrieved May 15, 2015, from http://nlp.stanford.edu/~socherr/SocherLeManningNg_nipsDeepWorkshop2013.pdf

Socher, R., & Lin, C. (2011). Parsing natural scenes and natural language with recursive neural networks. *Proceedings of the ...*, 129–136. Retrieved May 15, 2015, from http://machinelearning.wustl.edu/mlpapers/paper_files/ICML2011Socher_125.pdf

Soyata, T., Ba, H., Heinzelman, W., Kwon, M., & Shi, J. (2014). Accelerating Mobile-Cloud Computing. In *Communication Infrastructures for Cloud Computing* (pp. 175–197). doi:10.4018/978-1-4666-4522-6.ch008

Sundaresan, S., Teixeira, R., Tech, G., Feamster, N., Pescapè, A., & Crawford, S. (2011). Broadband Internet Performance : A View From the Gateway. *ACM SIGCOMM Computer Communication Review*, *41*(4), 134–145.

The Apache Software Foundation. (2014a). Apache Hadoop Project. *Hadoop*. Retrieved February 18, 2014, from http://hadoop.apache.org

The Apache Software Foundation. (2014b). Apache Mahout Project. *Mahout*. Retrieved July 18, 2014, from https://mahout.apache.org

The Apache Software Foundation. (2014c). Apache Cassandra Database. *Cassandra*. Retrieved February 18, 2014, from http://cassandra.apache.org/

The Open Networking Foundation. (2013). OpenFlow Switch Specification, Version 1.4.0. Retrieved April 10, 2014, from https://www.opennetworking.org/images/stories/downloads/sdn-resources/onf-specifications/openflow/openflow-spec-v1.4.0.pdf

Vissicchio, S., Vanbever, L., & Bonaventure, O. (2014). Opportunities and research challenges of hybrid software defined networks. *ACM SIGCOMM Computer Communication Review*, *44*(2), 70–75. Retrieved May 15, 2015, from http://dl.acm.org/citation.cfm?doid=2602204.2602216

Waldrop, M. (2008). Big data: Wikiomics. Nature 455, 22-25.

Wang, H., Liu, W., & Soyata, T. (2014). Accessing Big Data in the Cloud Using Mobile Devices. In P. I. S. R. Hershey (Ed.), *Handbook of Research on Cloud Infrastructures for Big Data Analytics* (pp. 444–470). doi:10.4018/978-1-4666-5864-6.ch018

Ward, J. S., & Barker, A. (2013). Undefined By Data: A Survey of Big Data Definitions. *arXiv.org*. Retrieved May 15, 2015, from http://arxiv.org/abs/1309.5821\npapers3://publication/uuid/63831F5F-B214-46D5-8A86-671042BE993F

Yiakoumis, Y., Yap, K.-K., Katti, S., Parulkar, G., & McKeown, N. (2011). Slicing home networks. In *Proceedings of the 2nd ACM SIGCOMM workshop on Home networks - HomeNets '11* (pp. 1-6).

Zhou, R., & Li, T. (2013). Leveraging Phase Change Memory to Achieve Efficient Virtual Machine Execution. In International Conference on Virtual Execution Environments (*VEE)* (pp. 179-190).